\documentclass[aps,12pt,preprint,groupedaddress,floatfix]{revtex4}
\usepackage{graphicx}
\usepackage{dcolumn}
\usepackage{bm}
\usepackage{amssymb}
\usepackage{amsmath}
\usepackage{color}

\begin{document}

\title{Microscopic study of induced fission dynamics of $^{226}$Th with covariant energy density functionals}
\author{H. Tao$^{1}$}
\author{J. Zhao$^{2,3}$}
\author{Z. P. Li$^{1}$}\email{zpliphy@swu.edu.cn}
\author{T. Nik\v si\'c$^{3}$}
\author{D. Vretenar$^{3}$}

\affiliation{$^{1}$School of Physical Science and Technology, Southwest University, Chongqing 400715, China}
\affiliation{$^{2}$Microsystem \& Terahertz Research Center, China Academy of Engineering Physics, Chengdu 610200, Sichuan, China}
\affiliation{$^{3}$Physics Department, Faculty of Science, University of Zagreb, 10000 Zagreb, Croatia}

\bigskip
\date{\today}

\begin{abstract}
Static and dynamic aspects of the fission process of $^{226}$Th are analyzed in a self-consistent framework based on
relativistic energy density functionals. Constrained relativistic mean-field (RMF) calculations in the collective space of
axially symmetric quadrupole and octupole deformations, based on the energy density functional PC-PK1 and a
$\delta$-force pairing, are performed to determine the potential energy surface of the fissioning nucleus, the scission line,
the single-nucleon wave functions, energies and occupation probabilities, as functions of deformation parameters. Induced fission dynamics is described using the
time-dependent generator coordinate method in the Gaussian overlap approximation. A collective Schr\"odinger equation,
determined entirely by the microscopic single-nucleon degrees of freedom, propagates adiabatically in time the initial wave packet
built by boosting the ground-state solution of the collective Hamiltonian for  $^{226}$Th. The position of the scission line and the
microscopic input for the collective Hamiltonian are analyzed as functions of the strength of the pairing interaction. The effect
of static pairing correlations on the pre-neutron emission charge yields and total kinetic energy of fission fragments is examined in
comparison with available data, and the distribution of fission fragments is analyzed for different values of the initial excitation energy.
\end{abstract}


\maketitle

\section{\label{secI}Introduction}
A microscopic description of fission presents one of the most complex problems in low-energy theoretical nuclear physics \cite{Krappe12,Schunck16}.
For a comprehensive recent review and an exhaustive list of references, we refer the reader to Ref.~\cite{Schunck16}. The spontaneous or induced
fission process in which a heavy nucleus splits into fragments is out of reach for {\em ab initio} methods and, therefore, modern microscopic approaches are based on the framework of nuclear energy density functionals (NEDFs). Nuclear density functional theory (DFT) and its time-dependent (TD) generalization have enabled a self-consistent treatment of both static and dynamic aspects of fission \cite{Scamps15,Simenel14,Goddard15,Avez08,Bulgac16,Godd16,Tani17,Naka16,Giul14}. The slow large-amplitude collective motion of the compound system that eventually leads to the formation of the final fragments can be described, in a first approximation, as an adiabatic process in which the intrinsic nucleonic degrees of freedom are decoupled from macroscopic collective degrees of freedom such as multipole moments (deformations) of the mass distribution and pairing fields \cite{Schunck16}.

Numerous studies of spontaneous fission, based on NEDFs, have analyzed the effects of the choice of collective coordinates (shape degrees of freedom), approximations used to calculate the collective inertia, and coupling between shape and pairing degrees of freedom on fission half-lives \cite{Chinn92,Stas13,Warda12,Negele82,Sadhukhan13,Sadhu14,Smol95}. A quantitative description of induced fission is, in this framework, conceptually and computationally more challenging and this process has been explored less systematically. In particular, several recent studies have used the time-dependent generator coordinate method (TDGCM) \cite{Berger91} to compute the induced fission fragment charge and mass distributions \cite{Goutte05,Younes12,Regnier16c,Regnier16,Zdeb17}. In this approach the nuclear wave function is described as a linear superposition of many-body functions parametrized by a vector of collective coordinates. Assuming that the norm kernels of these many-body functions can be approximated by a Gaussian form factor (Gaussian overlap approximation GOA), the GCM Hill-Wheeler equation reduces to a local, time-dependent, Schr\"odinger-like equation in the space of collective coordinates. In this approach the dynamics of the fissioning system essentially depends on the choice of the collective coordinates, energy density functional, pairing interaction, and approximations used to calculate the effective inertia \cite{Regnier16}.

Applications of the TDGCM+GOA have so far been based on non-relativistic Skyrme and Gogny functionals. Relativistic functionals, equally successful in mean-field and beyond mean-field (GCM) nuclear structure applications \cite{VALR.05,Meng06,Stone07,Meng16}, have only been employed in analyses of fission barriers and spontaneous fission \cite{Li10a,BNL12,BNL14,Jieb15,Jie15,Jie16,Burvenich2004_PRC69-014307,Blum1994_PLB323-262,Zhang2003_CPL20-1694,Bender2003_RMP75-121,Lu2006_CPL23-2940,Abusara2010_PRC82-044303,Abusara2012_PRC85-024314,Agbemava2017_PRC95-054324,Zhou2016_PS91-063008,Karatzikos2010_PLB689-72}. Several recent studies have performed multidimensionally constrained self-consistent relativistic mean-field calculations of deformation energy surfaces and fission barriers  of actinide nuclei \cite{Abusara2010_PRC82-044303,Karatzikos2010_PLB689-72,BNL14,Jieb15} and superheavy nuclei \cite{Karatzikos2010_PLB689-72,Abusara2012_PRC85-024314,Agbemava2017_PRC95-054324}. We have also analyzed the effects of triaxial and octupole deformations \cite{Jie15}, and the coupling between shape and pairing degrees of freedom \cite{Jie16} on dynamic spontaneous fission paths and half-lives.
In this work we extend our approach and apply the framework of relativistic EDFs and the corresponding collective Hamiltonian
to an analysis of induced fission dynamics, making use of a recent implementation of the TDGCM+GOA \cite{Regnier16c}.
Section \ref{secII} presents an outline of the model used to calculate the potential energy surface, collective inertia, and the time evolution of the fissioning system. An illustrative calculation of induced fission of $^{226}$Th, for which the charge distribution of fission fragments displays symmetric and asymmetric peaks, is discussed in Sec.~\ref{secIII}. In particular, we study the sensitivity of pre-neutron emission charge yields and total kinetic energy of fission fragments on static pairing correlations. Section \ref{secIV} contains a summary of results and an outlook for future studies.

\section{\label{secII} Theoretical framework}
\subsection{Time-dependent Schr\"odinger-like equation for fission dynamics}
Nuclear fission can be modeled as a slow adiabatic process determined by only a few collective degrees of freedom. In the present study we consider the axial deformation parameters: quadrupole $\beta_2$ and octupole $\beta_3$. A time-dependent Schr\"odinger-like equation describes low-energy fission dynamics, and this equation can be derived using the time-dependent generator coordinate method (TDGCM) in the Gaussian overlap approximation (GOA) \cite{Regnier16,Schunck16}:
\begin{equation}
\label{TDSch}
i\hbar\frac{\partial}{\partial t}g(\beta_2, \beta_3, t)=\left[-\frac{\hbar^2}{2}\sum\limits_{kl}\frac{\partial}{\partial \beta_k}B_{kl}(\beta_2, \beta_3)\frac{\partial}{\partial \beta_l}+V(\beta_2, \beta_3)\right]g(\beta_2, \beta_3, t) \; ,\\
\end{equation}
where $g(\beta_2, \beta_3, t)$ denotes the complex wave function of the collective variables $(\beta_2, \beta_3)$ and time $t$. $V(\beta_2, \beta_3)$ and $B_{kl}(\beta_2, \beta_3)$ are the collective potential and mass tensor, respectively, and they completely determine the
dynamics of the fission process in the
TDGCM+GOA framework. These quantities will here be calculated in a self-consistent mean-field approach based on relativistic energy density functionals, as detailed in Sec. \ref{ssecII-II}. For the time-evolution we follow the method of Refs. \cite{Regnier16,Regnier16c} and make use of the software package FELIX \cite{Regnier16c} that solves the equations of the TDGCM in N-dimensions under the Gaussian overlap approximation.

From the Schr\"odinger-like equation (\ref{TDSch}) a continuity equation for the probability density $|g(\beta_2, \beta_3, t)|^2$ is obtained,
\begin{equation}
\label{continuity}
\frac{\partial}{\partial t}|g(\beta_2, \beta_3, t)|^2=-\nabla\cdot\mathbf{J}(\beta_2, \beta_3, t) \; ,
\end{equation}
where $\mathbf{J}(\beta_2, \beta_3, t)$ is the probability current defined by the relation:
\begin{equation}
\label{current}
J_k(\beta_2, \beta_3, t)=\frac{\hbar}{2i}\sum\limits_{l=2}^3 B_{kl}(\beta_2, \beta_3)\left[g^*(\beta_2, \beta_3,t)\frac{\partial g(\beta_2, \beta_3,t)}{\partial\beta_l}- g(\beta_2, \beta_3,t)\frac{\partial g^*(\beta_2, \beta_3,t)}{\partial\beta_l}\right] \; .
\end{equation}

The collective space is divided into the inner region in which the nuclear density distribution is whole, and an external region that contains the two fission fragments. The set of scission configurations defines the hyper-surface that separates the two regions. The flux of the probability current through this
hyper-surface provides a measure of the probability of observing a given pair of fragments at time $t$.
For a surface element $\xi$ on the scission hyper-surface, the integrated flux $F(\xi, t)$ is defined as  \cite{Regnier16c}:
\begin{equation}
\label{eq:Ft}
F(\xi, t)=\int_{t=0}^t dt\int_{(\beta_2,\beta_3)\in\xi}\mathbf{J}(\beta_2,\beta_3,t)\cdot d\mathbf{S}.
\end{equation}
For each scission point, $(A_L, A_H)$ denote the masses of the lighter and heavier fragments, respectively.  Therefore, the yield for the fission fragment with mass $A$ can be defined by
\begin{equation}
Y(A)\propto\sum\limits_{\xi\in{\cal A}}\lim\limits_{t\to+\infty}F(\xi, t),
\end{equation}
where ${\cal A}$ is the set of all elements $\xi$ belonging to the scission hyper-surface such that one of the fragments has mass $A$.

\subsection{\label{ssecII-II}Collective parameters}

The entire dynamics of the Schr\"odinger-like equation (\ref{TDSch}) is governed by the four functions of the intrinsic deformations $\beta_2$ and $\beta_3$: the collective potential $V$ and the three mass parameters $B_{22}$, $B_{23}$, $B_{33}$. These functions are determined by performing
constrained self-consistent mean-field calculations for a specific choice of the nuclear energy density functional and pairing interaction.
In the present study the energy density functional PC-PK1 \cite{Zhao10} determines the effective interaction in the particle-hole channel, and a $\delta$-force is used in the particle-particle channel.

The entire map of the energy surface as function of the quadrupole and octupole
deformations is obtained by imposing constraints on the quadrupole and octupole mass moments.
The method of quadratic constraints uses an unrestricted variation of the function
\begin{equation}
\langle H\rangle
   +\sum_{k=2,3}{C_{k}\left(\langle \hat{Q}_{k}  \rangle - q_{k}  \right)^2} \; ,
\label{constr}
\end{equation}
where $\langle H\rangle$ is the total energy, and  $\langle \hat{Q}_{k}\rangle$
denotes the expectation value of the mass quadrupole and octupole operators:
\begin{equation}
\hat{Q}_{2}=2z^2-r_\bot^2 \quad \textnormal{and}\quad \hat{Q}_{3}=2z^3-3zr_\bot^2 \;.
\end{equation}
$q_{k}$ is the constrained value of the multipole moment,
and $C_{k}$ is the corresponding stiffness constant~\cite{RS.80}.
The corresponding deformation parameters $\beta_2$ and $\beta_3$ can be determined from the following relations:
\begin{eqnarray}
\beta_2 &=& \frac{\sqrt{5\pi}}{3AR_0^2}\langle \hat Q_2\rangle , \\
\beta_3 &=& \frac{\sqrt{7\pi}}{3AR_0^3}\langle \hat Q_3\rangle ,
\end{eqnarray}
with $R_0=r_0A^{1/3}$ and $r_0=1.2$ fm.

The single-nucleon wave functions, energies and occupation factors,
generated from constrained self-consistent solutions of the relativistic mean-field  plus BCS-pairing
equations (RMF+BCS), provide the microscopic input for the parameters of the
Schr\"odinger-like equation (\ref{TDSch}).
The solution of the single-nucleon Dirac equation is obtained by expanding the nucleon wave functions in an axially
deformed harmonic oscillator basis, as described in appendix \ref{app-A}.

The mass tensor associated with $q_2=\langle\hat{Q}_{2}\rangle$ and $q_3=\langle\hat{Q}_{3}\rangle$ are calculated in the
perturbative cranking approximation~\cite{Girod79,ZPL16}
\begin{equation}
\label{eq:BB}
B_{kl}(q_2,q_3)=\frac{2}{\hbar^2}
 \left[\mathcal{M}_{(1)} \mathcal{M}^{-1}_{(3)} \mathcal{M}_{(1)}\right]_{kl}\;,
\end{equation}
with
\begin{equation}
\label{masspar-M}
\mathcal{M}_{(n),kl}(q_2,q_3)=\sum_{i,j}
 {\frac{\left\langle i\right|\hat{Q}_{k}\left| j\right\rangle
 \left\langle j\right|\hat{Q}_{l}\left| i\right\rangle}
 {(E_i+E_j)^n}\left(u_i v_j+ v_i u_j \right)^2}\;.
\end{equation}
The summation is over the proton and neutron quasiparticle states.
The quasiparticle energies $E_i$, occupation probabilities $v_i$, and
single-nucleon states are determined by solutions of the constrained RMF+BCS equations.

The collective energy surface includes the energy of zero-point motion, which has to be
subtracted. The vibrational and rotational zero-point energy (ZPE) corrections are calculated
in the cranking approximation \cite{Ni.09,Li09a}:
\begin{equation}
\label{ZPE-vib}
\Delta V_{\rm vib}(\beta_2, \beta_3) = \frac{1}{4}
\textnormal{Tr}\left[\mathcal{M}_{(3)}^{-1}\mathcal{M}_{(2)}  \right]\;,
\end{equation}
and
\begin{equation}
\label{ZPE-rot}
\Delta V_{\rm rot}(\beta_2, \beta_3)=\frac{\langle\hat J^2\rangle}{2{\cal I}}\;,
\end{equation}
respectively, where ${\cal I}$ is the Inglis-Belyaev moment of inertia \cite{Inglis56,Belyaev61}.
The potential $V(\beta_2, \beta_3)$ in the time-dependent collective equation (\ref{TDSch}) is
obtained by subtracting the ZPE corrections from the total mean-field energy:
\begin{equation}
\label{eq:Vcoll}
V(\beta_2, \beta_3)=E_{\rm tot}(\beta_2, \beta_3)-\Delta V_{\rm vib}(\beta_2, \beta_3)-\Delta V_{\rm rot}(\beta_2, \beta_3).
\end{equation}

\section{\label{secIII} Results and discussion}

In this section we present the results of an illustrative study of induced fission of $^{226}$Th, for which the charge distribution of fission fragments exhibits a coexistence of symmetric and asymmetric peaks \cite{Schmidt01}. In the first step a large-scale deformation-constrained self-consistent RMF+BCS calculation is performed to generate the potential energy surface and single-nucleon wave functions in the $(\beta_2, \beta_3)$ plane. The range of collective variables is -0.83 to 6.01 for $\beta_2$ with a step $\Delta\beta=0.04$, and from 0.01 to 3.53 for $\beta_3$ with a step $\Delta\beta_3=0.08$. The energy density functional PC-PK1 \cite{Zhao10} is used for the effective interaction in the particle-hole channel, and a $\delta$-force pairing with strengths parameters: $V_n=360$ MeV~fm$^{3}$ and $V_p=378$ MeV~fm$^{3}$ determined by the empirical pairing gap parameters of $^{226}$Th,
calculated using a five-point formula \cite{Bender00}. The self-consistent Dirac equation for the single-particle wave functions is solved by expanding the nucleon spinors in an axially deformed harmonic oscillator basis in cylindrical coordinates with 20 major shells. The computer code FELIX \cite{Regnier16c} is used for modelling the time-evolution of the fissioning nucleus with a time step $\delta t=5\times 10^{-4}$ zs. The parameters of the additional imaginary absorption potential that takes into account the escape of the collective wave packet  in the domain outside the region of calculation \cite{Regnier16c} are: the absorption rate $r=20\times 10^{22}$ s$^{-1}$, and the width of the absorption band $w=1.5$.

\subsection{\label{ssecI} Potential energy surface, scission line, and total kinetic energy}

\begin{figure}[htb]
\includegraphics[scale=0.49]{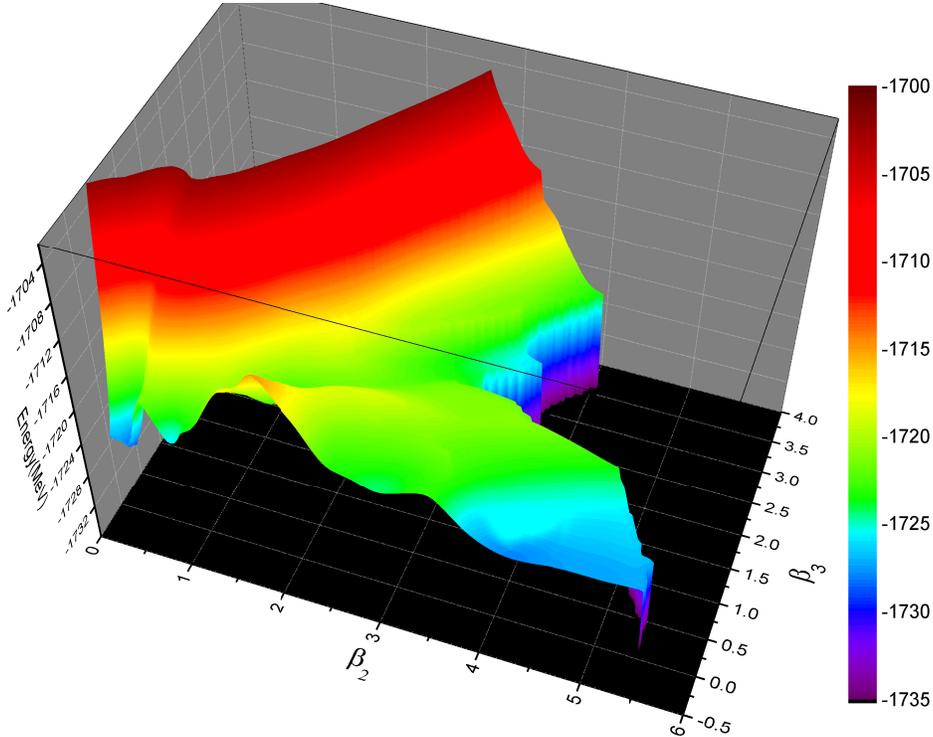}
\caption{\label{fig:3DPES} (Color online) Self-consistent RMF+BCS
quadrupole and octupole constrained deformation energy surface (in MeV) of $^{226}$Th in the $\beta_2-\beta_3$ plane.}
\end{figure}

\begin{figure}[htb]
\includegraphics[scale=0.5]{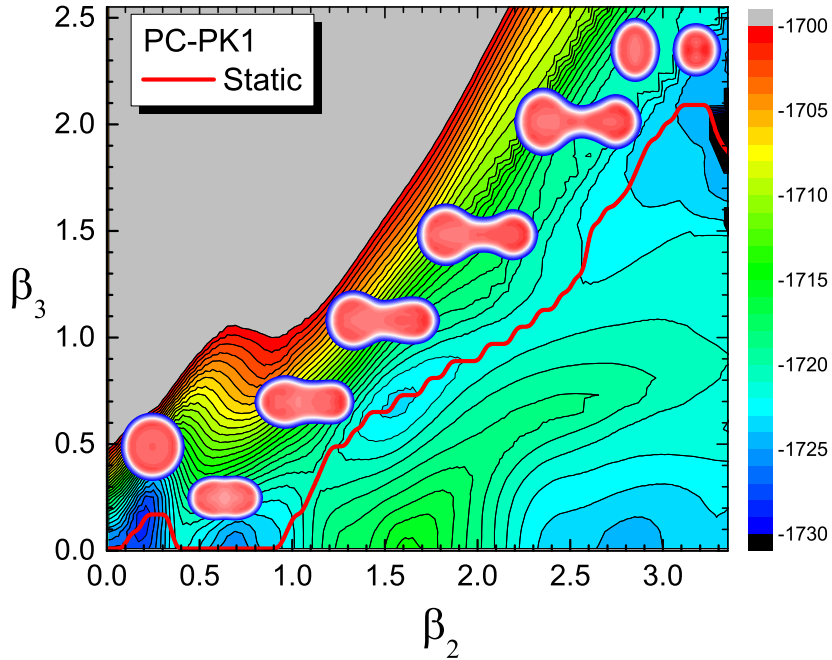}
\caption{\label{fig:PES2} (Color online) Same as in the caption to Fig.~\ref{fig:3DPES}, but plotted as a countour map.
The red curve is the static fission path and the density distributions for selected deformations
along the fission path are also shown.}
\end{figure}

The present RMF+BCS results for the potential energy surface (PES), scission line, and total kinetic energy of $^{226}$Th can be compared to those obtained in Ref. \cite{Dubray08} using the Hartree-Fock-Bogoliubov framework based on the Gogny D1S effective interaction.
Figures~\ref{fig:3DPES} and \ref{fig:PES2} display the self-consistent RMF+BCS quadrupole and octupole constrained energy surfaces, the static fission path, and density distributions for selected deformations along the fission path of $^{226}$Th. The lowest minimum is located at $(\beta_2, \beta_3)\sim(0.20,0.17)$, but is rather soft against octupole deformation. A triple-humped fission barrier is predicted along the static fission path, and the calculated heights are 7.10, 8.58, and 7.32 MeV from the inner to the outer barrier, respectively. At elongations $\beta_2>1.5$ a symmetric valley extends up to the scission point at  $\beta_2\sim5.4$. The symmetric and asymmetric fission valleys are separated by a ridge from $(\beta_2, \beta_3)=(1.6, 0.0)$ to (3.4, 1.0). One notices that the overall topography of the PES is similar to that calculated with the Gogny D1S interaction \cite{Dubray08}.


\begin{figure}[htb]
\includegraphics[scale=0.4]{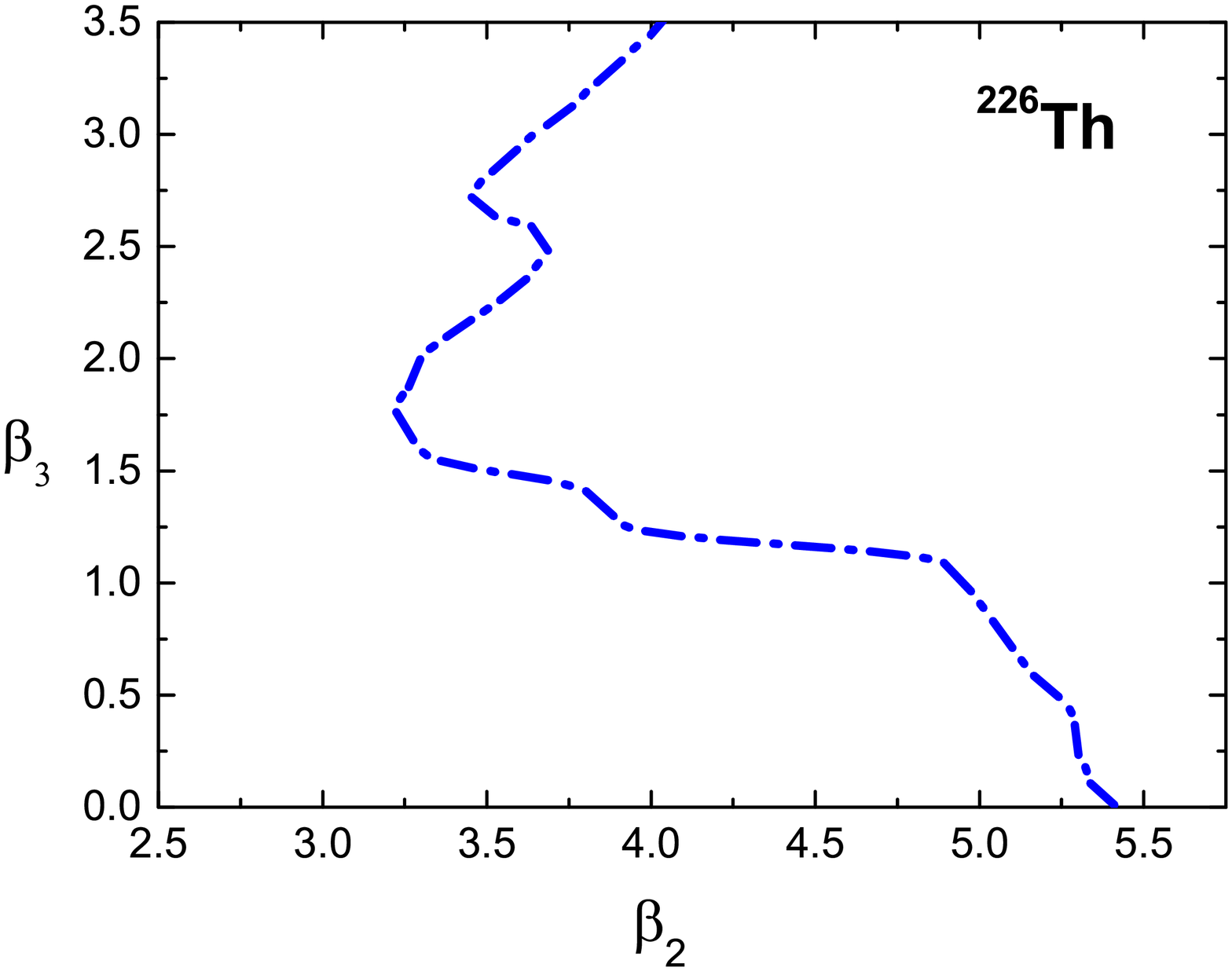}
\caption{\label{fig:scission1}(Color online) The scission contour of $^{226}$Th in the $\beta_2-\beta_3$ plane.}
\label{scission1}
\end{figure}
When describing fission in the $\beta_2-\beta_3$ collective space, scission is characterized by a discontinuity between the two domains of pre- and postscissioned configurations. Scission can be described using the Gaussian neck operator $\hat Q_N=\exp\left[-(z-z_N)^2/a^2_N\right]$, where $a_N=1$ fm and $z_N$ is the position of the neck \cite{Younes09}. It is related to the number of particles in the neck, and here we follow the prescription of Ref. \cite{Regnier16} to define the pre-scission domain by $\langle\hat Q_N\rangle>3$ and consider the frontier of this domain as the scission line. In Fig.~\ref{scission1} we plot the scission profile for $^{226}$Th in the $\beta_2-\beta_3$ plane. The curve starts from an elongated symmetric point at $\beta_2\sim5.4$ and evolves to a minimal  elongation with $\beta_2\sim3.2$ as asymmetry increases. From that point $\beta_3$ increases rapidly along the scission line and we also note a more gradual increase of the quadrupole deformation parameter. The general pattern is similar to the scission lines for $^{226}$Th obtained in Refs. \cite{Dubray08,Younes09}.

\begin{figure}[htb]
\includegraphics[scale=0.4]{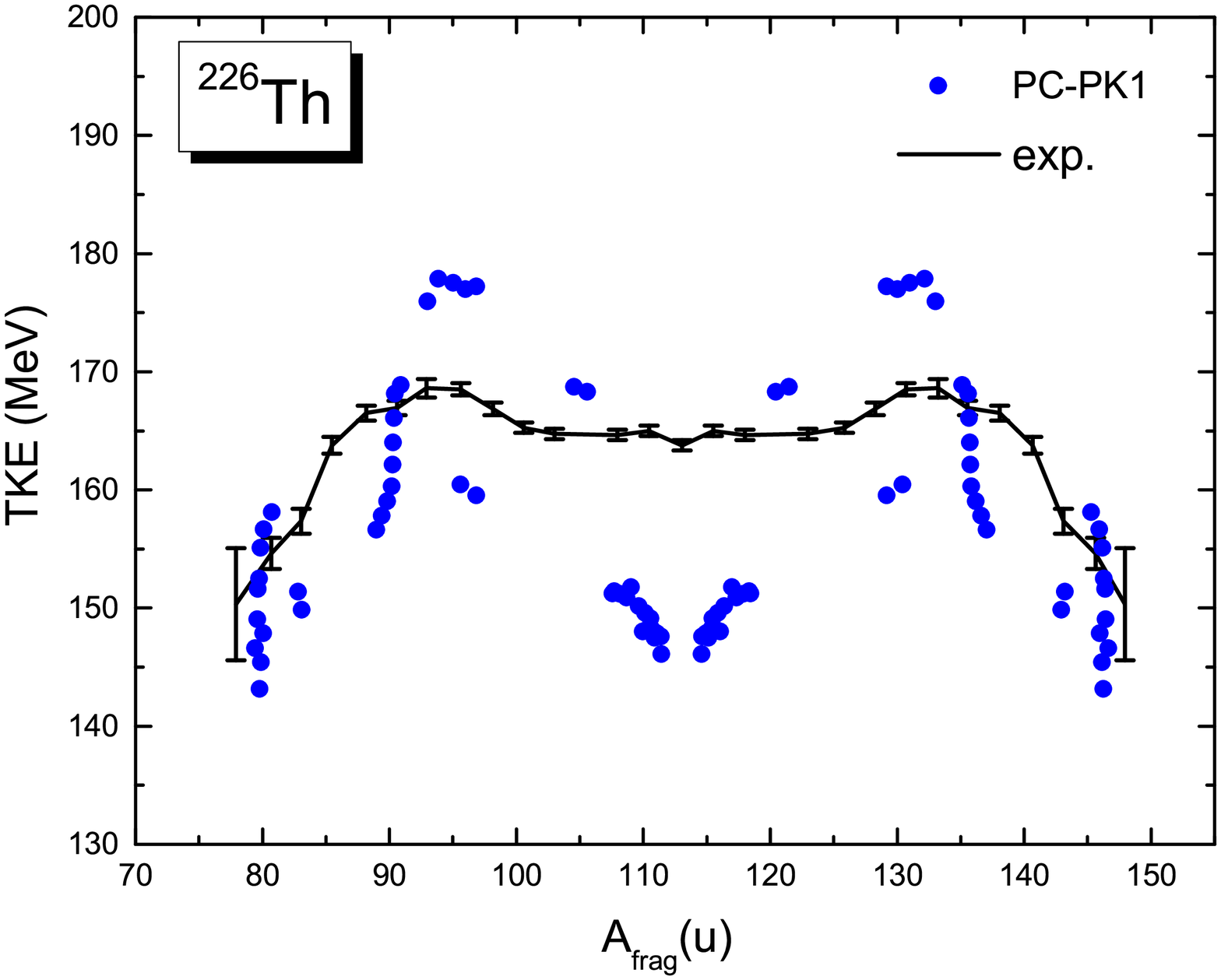}
\caption{\label{fig:TKE1}(Color online) The calculated total kinetic energy of the nascent fission fragments for $^{226}$Th as a function of fragment mass, in comparison to the data \cite{Schmidt01}.}
\end{figure}

The total kinetic energy (TKE) for a particular pair of fragments can be evaluated from
\begin{equation}
E_{\rm TKE}=\frac{e^2Z_HZ_L}{d_{\rm ch}},
\label{eq:TKE}
\end{equation}
where $e$ is the proton charge, $Z_H (Z_L)$ the charge of the heavy (light) fragment, and $d_{\rm ch}$ the distance between fragment centers of charge at scission. Figure \ref{fig:TKE1} displays the calculated total kinetic energies of the nascent fission fragments for $^{226}$Th as a function of fragment mass. For comparison, the data obtained in photo-induced fission measurement \cite{Schmidt01} are also included in the figure.
One notices that the theoretical results qualitatively reproduce the trend of the data, in particular the maxima for $A_{\rm frag}\sim132$ and $A_{\rm frag}\sim94$. On a quantitative level the calculation exhibits more structure when compared to experiment. This may be due to the fact that the experimental values correspond to an excitation energy of the fissioning nucleus of the order of 11 MeV, whereas  formula (\ref{eq:TKE})  is valid only for low-energy fission. As it is well known, the kinetic energy distribution is generally smoothed out as the fission energy increases. In particular, the kinetic energy in the symmetric mass region increases \cite{Pomme94}, which explains why experimental TKEs display only a very shallow minimum for $A_{\rm frag} = A/2$. We note that the present theoretical results are consistent with those obtained using the Gogny D1S effective interaction in Ref.~\cite{Dubray08}.

\subsection{\label{ssecII} Sensitivity of the fission process to the choice of pairing strength}

\begin{figure}[htb]
\includegraphics[scale=0.5]{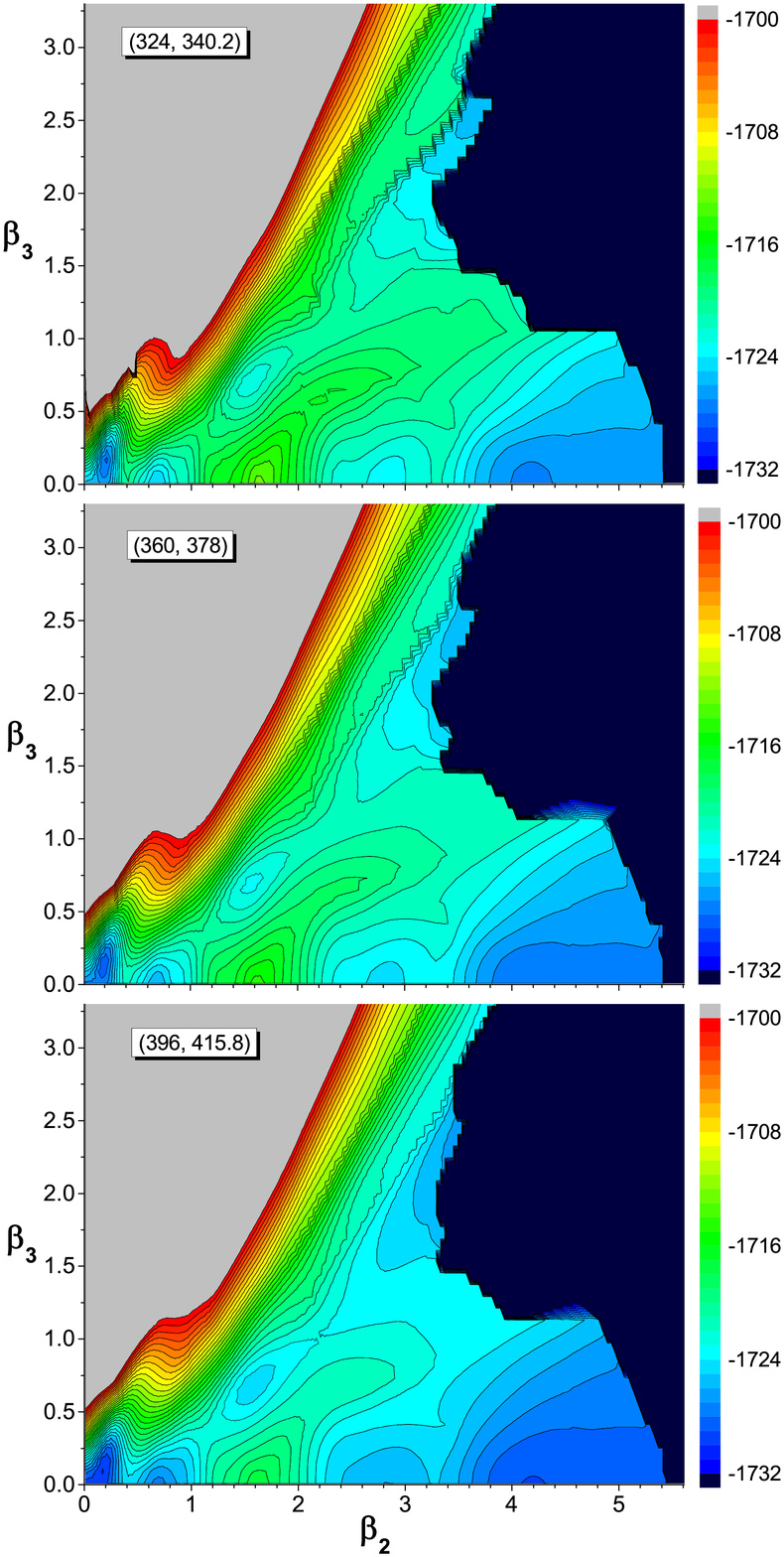}
\caption{\label{fig:PES3} (Color online) Potential energy surfaces of $^{226}$Th in the $\beta_2-\beta_3$ plane, calculated with the
functional PC-PK1 and for three parametrizations of the pairing force: $(V_n, V_p)=(324, 340.2)$ (top), (360, 378) (middle), and (396, 415.8) (bottom), in units of MeV$\cdot$fm$^{3}$.}
\end{figure}
\begin {table}[h]
\begin {center}
\caption{The height of the fission barriers (in MeV) with respect to the corresponding ground-state minima, for different
values of the pairing strengths.}
\bigskip
\begin {tabular}{cccccc}
\hline
\hline
 & $B_I$  & $B_{II}^{\rm asy}$ &  $B_{III}^{\rm asy}$  & $B_{II}^{\rm sym}$ &  $B_{III}^{\rm sym}$ \\
\hline
90\% pairing    & 8.23  &  9.47  & 7.74  & 15.64  &  6.38   \\
100\% pairing  & 7.10  &  8.58  & 7.32  & 14.21  &  5.72   \\
110\% pairing  & 5.92  &  7.78  & 7.09  & 12.72  &  5.17   \\
\hline
\end{tabular}
\label{tab:B}
\end{center}
\end{table}

\begin{figure}[htb]
\includegraphics[scale=0.5]{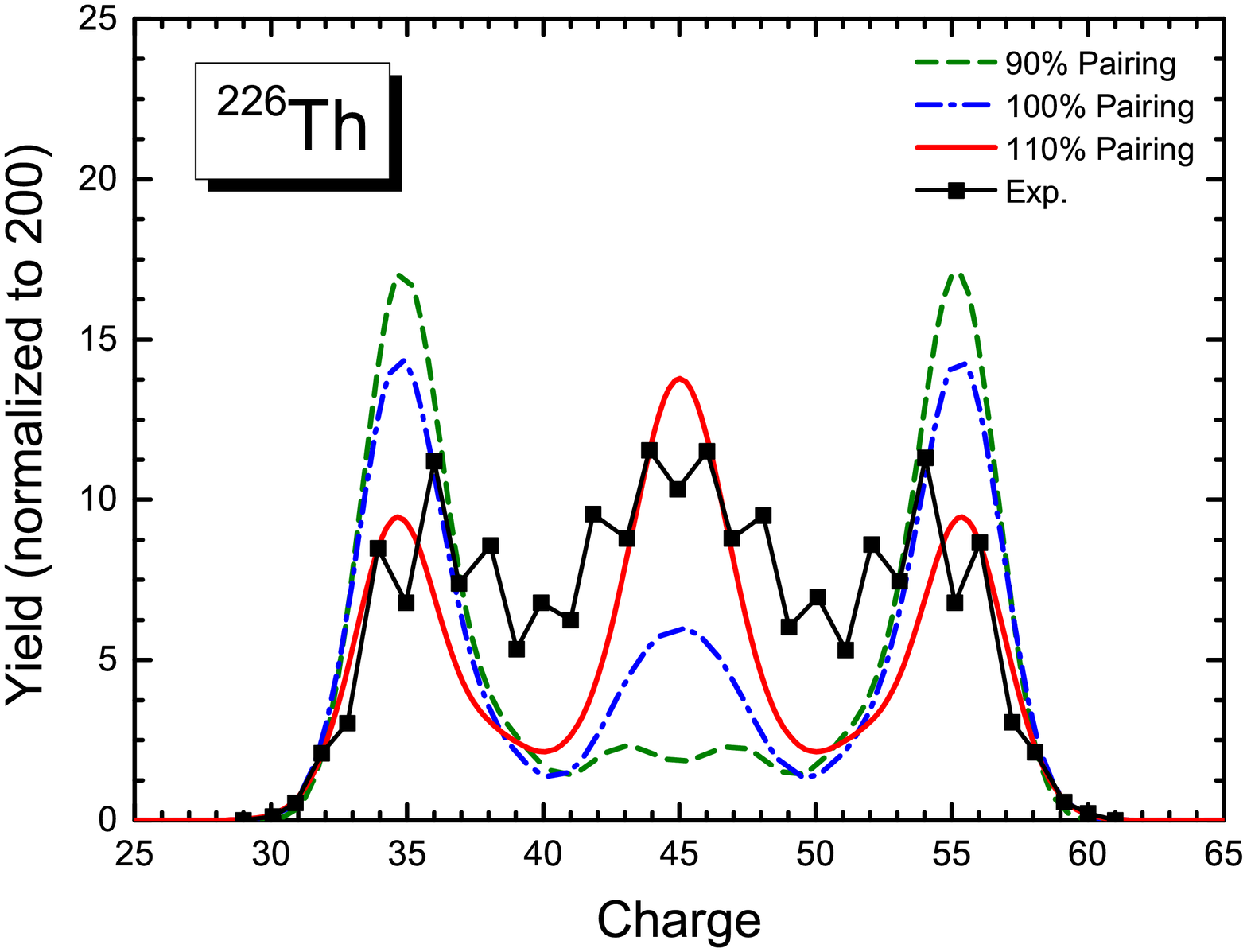}
\caption{\label{fig:Chargepair}  (Color online) Pre-neutron emission charge yields for photo-induced fission of $^{226}$Th. The results of calculations for three different values of the pairing strength are compared to the data \cite{Schmidt01}.}
\end{figure}

A number of model studies, including those based on the relativistic mean-field framework \cite{Abusara2010_PRC82-044303,Karatzikos2010_PLB689-72}, have shown that the height of calculated fission barriers is rather sensitive to the strength of pairing interaction. To illustrate the effect of pairing correlations on fission dynamics, we analyze the characteristics of the fission process for different strengths of the pairing interaction. Figure \ref{fig:PES3} displays the PESs of $^{226}$Th for three parametrizations of pairing force: $(V_n, V_p)=(324, 340.2)$, (360, 378), and (396, 415.8) MeV~fm$^{3}$. These values correspond to 90\%, 100\%, and 110\%, respectively, of the original pairing strengths that were determined to reproduce the empirical pairing gaps of $^{226}$Th. Even though the general topography of the PESs does not change significantly as pairing increases, the barriers are reduced considerably (see Table \ref{tab:B}). In particular, the ridge between the symmetric and asymmetric fission valleys is lowered, and this leads to pronounced competition between the two fission modes (c.f. Fig. \ref{fig:Chargepair}).

\begin{figure}[htb]
\includegraphics[scale=0.4]{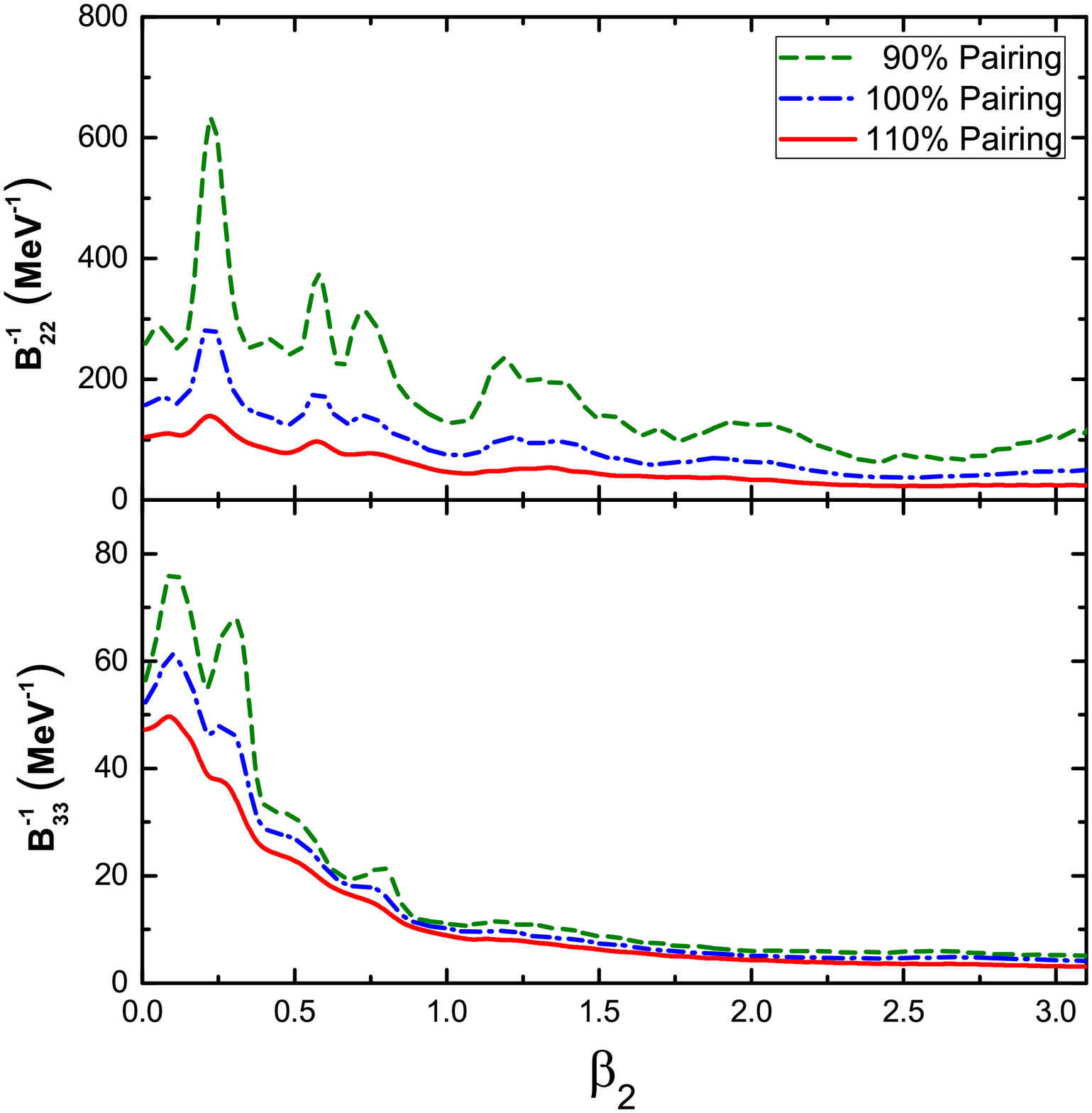}
\caption{(Color online)
Collective masses $B^{-1}_{22}$ and $B^{-1}_{33}$ related to vibrations in $\beta_2$ and $\beta_3$, respectively, along the static fission path for three values of the pairing strength.}
\label{fig:BBB}
\end{figure}
\begin{figure}[htb]
\includegraphics[scale=0.4]{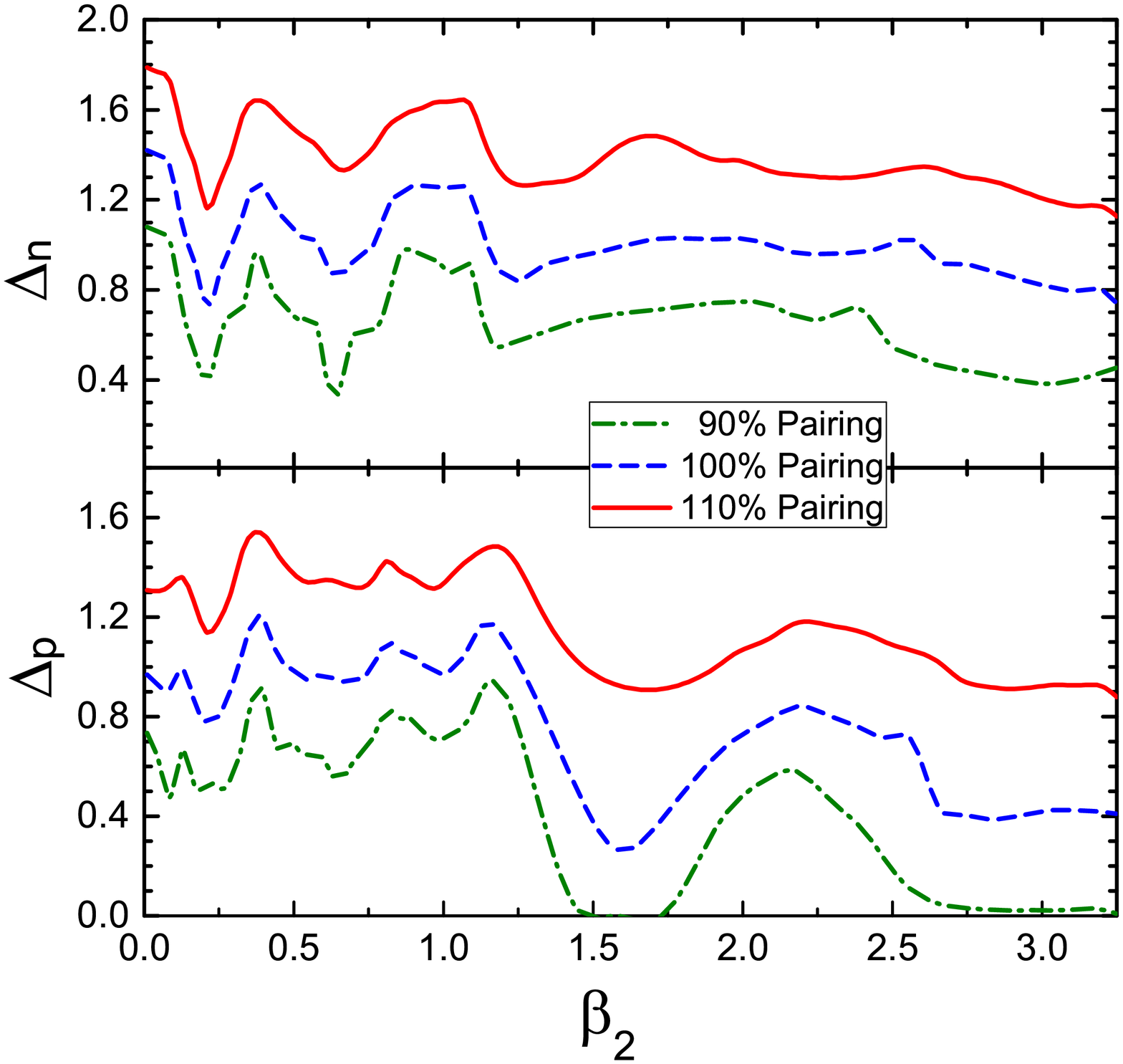}
\caption{\label{fig:Delta} (Color online) Pairing gaps for neutrons (upper panel) and protons (lower panel) along the static fission path.}
\end{figure}

In Fig. \ref{fig:BBB} we plot the collective masses $B^{-1}_{22}$ and $B^{-1}_{33}$, related to vibrations in $\beta_2$ and $\beta_3$, respectively, along the static fission path for the three choices of pairing strengths. They are elements of the inverse matrix of the mass tensor $B_{kl}$ in Eq. (\ref{eq:BB}). In general, the collective masses exhibit a rather complex behaviour for $\beta_2<1.0$, and show very little variation for large deformations. On the whole $B^{-1}_{22}$ gradually decreases as the nucleus is elongated, while $B^{-1}_{33}$ displays a pronounced decrease only in the region up to $\beta_2\sim1.0$. As pairing correlations increase, the collective masses are reduced and the shell oscillations are also smoothed out. These effects are illustrated in Fig. \ref{fig:Delta}, where we plot the neutron and proton pairing gaps along the static fission path for different pairing strengths. The  fluctuations of pairing gaps reflect the underlying shell structure, and pairing is strongly reduced wherever the level density around the Fermi level is small. As a result, the mass parameters are locally enhanced in regions of weak pairing.

\begin{figure}[htb]
\includegraphics[scale=0.45]{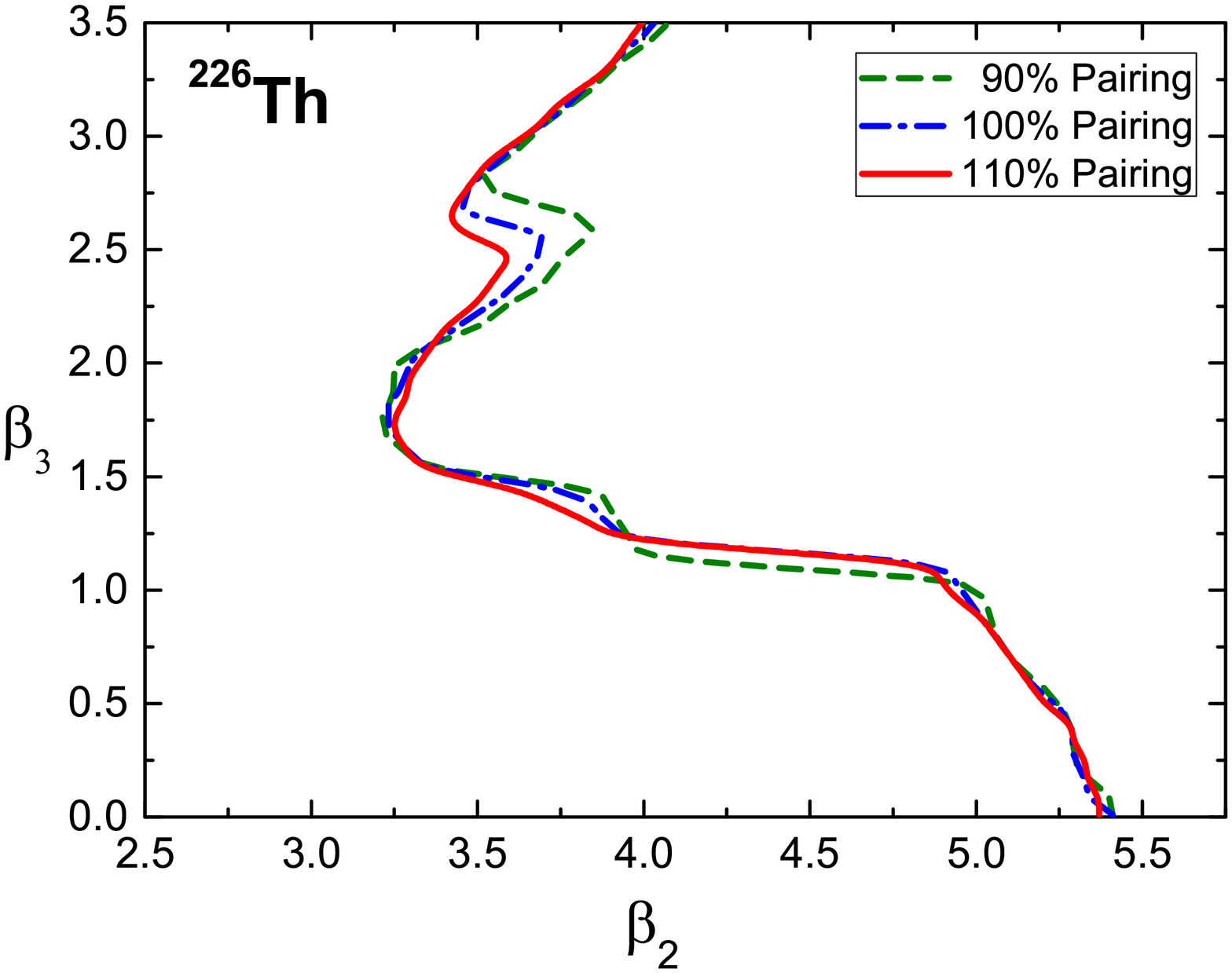}
\caption{(Color online)
The scission lines for $^{226}$Th in the $\beta_2-\beta_3$ plane, obtained in calculations with three different values of the pairing strength.}
\label{fig:scission}
\end{figure}
\begin{figure}[htb]
\includegraphics[scale=0.45]{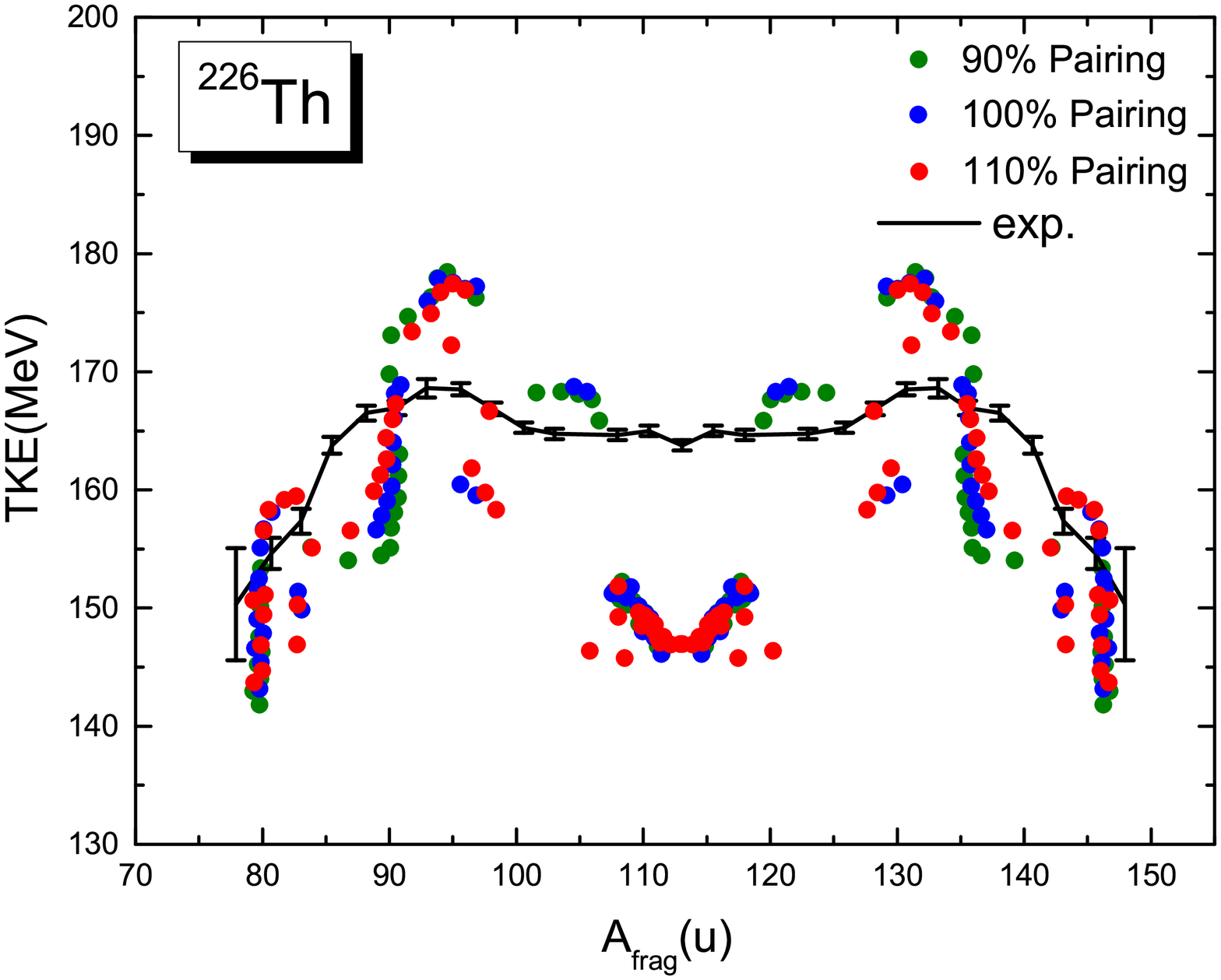}
\caption{(Color online)
Comparison between experimental and calculated total kinetic energy of nascent fission fragments for $^{226}$Th, as a function of
fragment mass and pairing strength.}
\label{fig:TKE3}
\end{figure}

Figures \ref{fig:scission} and \ref{fig:TKE3} display the scission lines in the $\beta_2-\beta_3$ plane and the TKEs of nascent fission fragments of $^{226}$Th, respectively, for three different values of the pairing strength. The pattern of the scission line does not change significantly, except at the
bending points and, overall, a smoother contour is obtained for stronger pairing. We also note that the scission points on the static fission path for three values of the pairing strength are very close to each other,  at $(\beta_2, \beta_3)\sim(3.3, 2.0)$. This result differs from that in $^{240}$Pu calculated using the HFB method with the Skyrme functional SkM$^*$ \cite{Schunck14}, where the quadrupole deformation $\beta_2$ at the scission point changes by as much as $\sim0.65$ when the original pairing strength is varied from 90\% to 110\%. Since the TKEs in the present study are fully determined by the scission configurations, varying the pairing strength does not lead to marked differences in the TKE distribution.

In Fig. \ref{fig:Chargepair} we compare the charge yields, obtained with three different pairing strengths, to the data for photo-induced fission of $^{226}$Th. Following the procedure of Ref. \cite{Regnier16}, the initial state is prepared by boosting the collective ground state in the
direction of increasing axial quadrupole deformation. The amplitude of the boost is determined so that
the average energy of the initial state is $\sim 1$ MeV above the corresponding asymmetric fission barrier $B_{II}^{\rm asy}$ of the collective potential energy surface [c.f. Eq. (\ref{eq:Vcoll})]. The calculation reproduces the trend of the data, except that obviously the model cannot describe the odd-even
staggering of the experimental charge yields.
For weak pairing correlations, that is, at 90\% of the original pairing strength, the yields are dominated by asymmetric fission with peaks at $Z=35$ and $Z=55$. A broad peak corresponding to symmetric fission is also predicted but is too low compared to data. This is because the asymmetric fission barrier  $B_{II}^{\rm asy}$ is $\sim6$ MeV lower than the symmetric one $B_{II}^{\rm sym}$. The asymmetric peaks are reduced and the symmetric peak enhanced as pairing correlations increase, and we find that the data are best reproduced by a pairing strength between 100\% and 110\% of the original parameters. This can be attributed to a reduction of the ridge between asymmetric and symmetric fission valleys when increasing the pairing strength. Another important effect is that the wavelength becomes longer because of smaller collective masses for stronger pairing, and this enhances the collective current in the symmetric fission valley beyond $\beta_2>2.5$.

\begin{figure}[htb]
\includegraphics[scale=0.5]{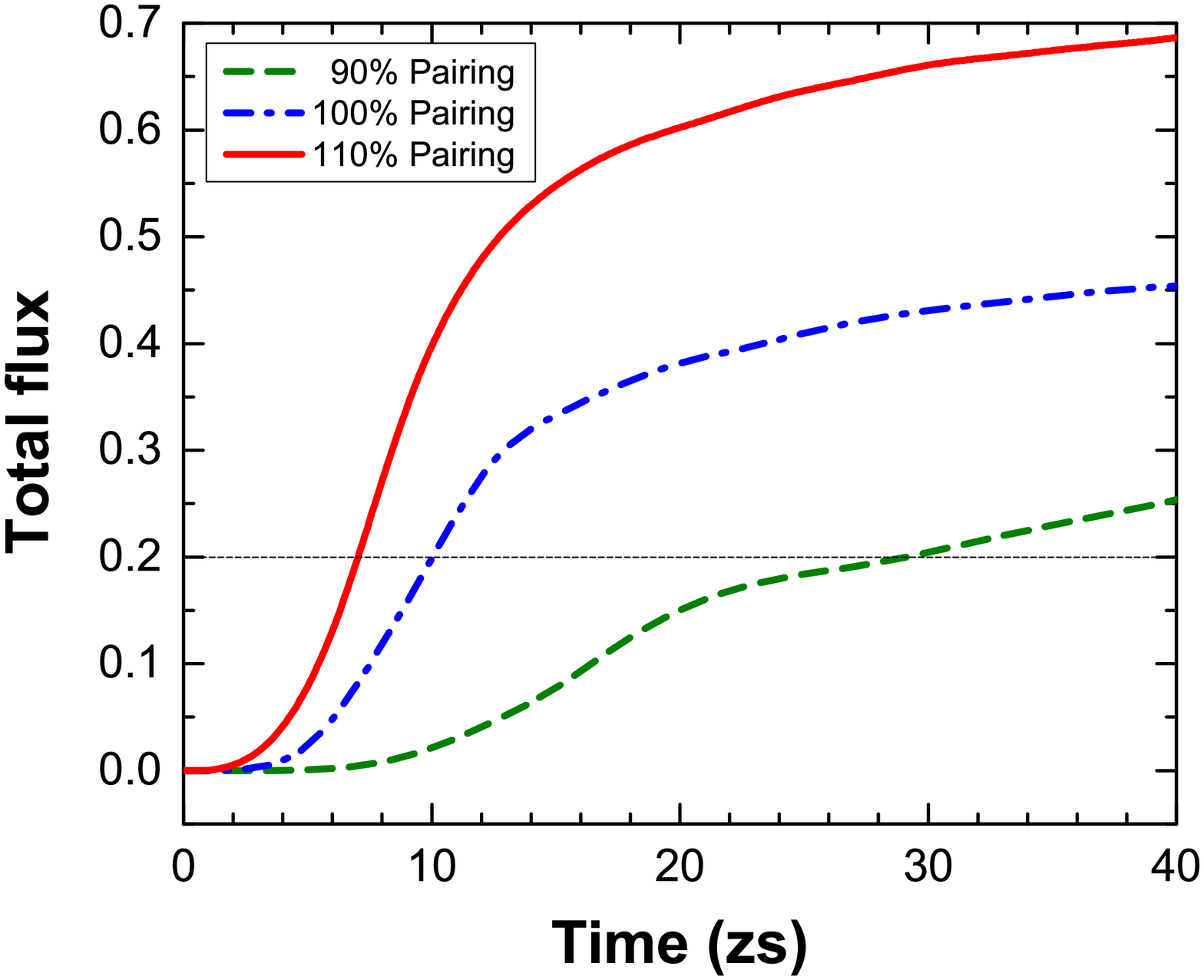}
\caption{\label{fig:Flux}  (Color online) Total flux as a function of time for three different pairing strengths. }
\end{figure}

Finally, we discuss the fission time for the nucleus by analyzing the total flux as a function of time in Fig. \ref{fig:Flux}. The total flux is obtained by integrating the flux $F(\xi, t)$ in Eq. (\ref{eq:Ft}) along the scission line. The fission time is obviously very sensitive to the pairing strength, and the time for the total flux to reach $1/5$ varies from $\sim30$ to $\sim7$ zs as the pairing strength changes from 90\% to 110\%. This is easy to understand because the current is proportional to the mass tensor $B_{kl}$ [c.f. Eq. (\ref{current})], which is enhanced for stronger pairing (c.f. Fig. \ref{fig:BBB}).

\subsection{\label{ssecIII} Sensitivity to the initial excitation energy}

\begin{figure}[htb]
\includegraphics[scale=0.5]{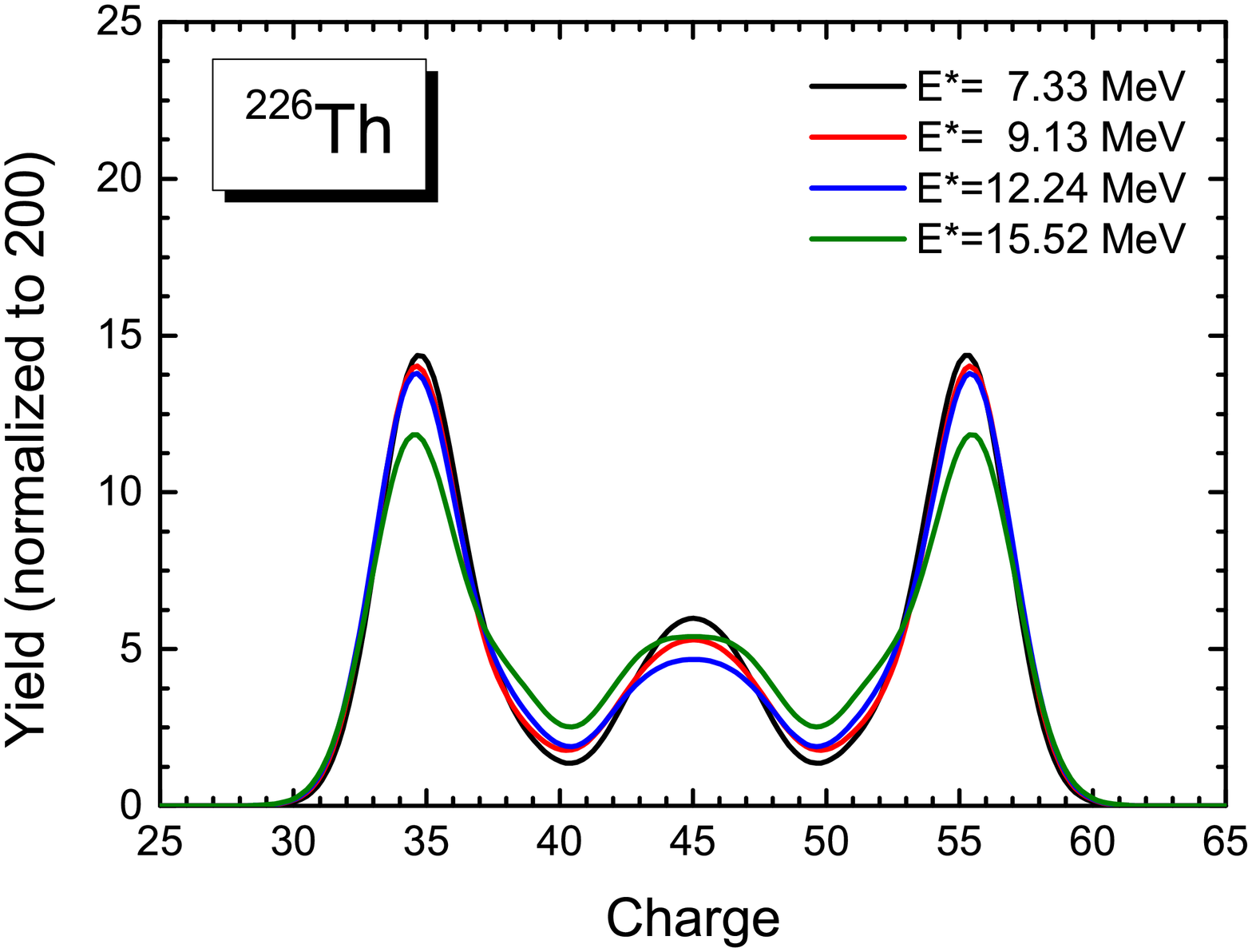}
\caption{\label{fig:ChargeEE}  (Color online)
Charge distribution of fission fragments for different excitation energies. The original pairing strength is used.}
\end{figure}
\begin{figure}[htb]
\includegraphics[scale=0.5]{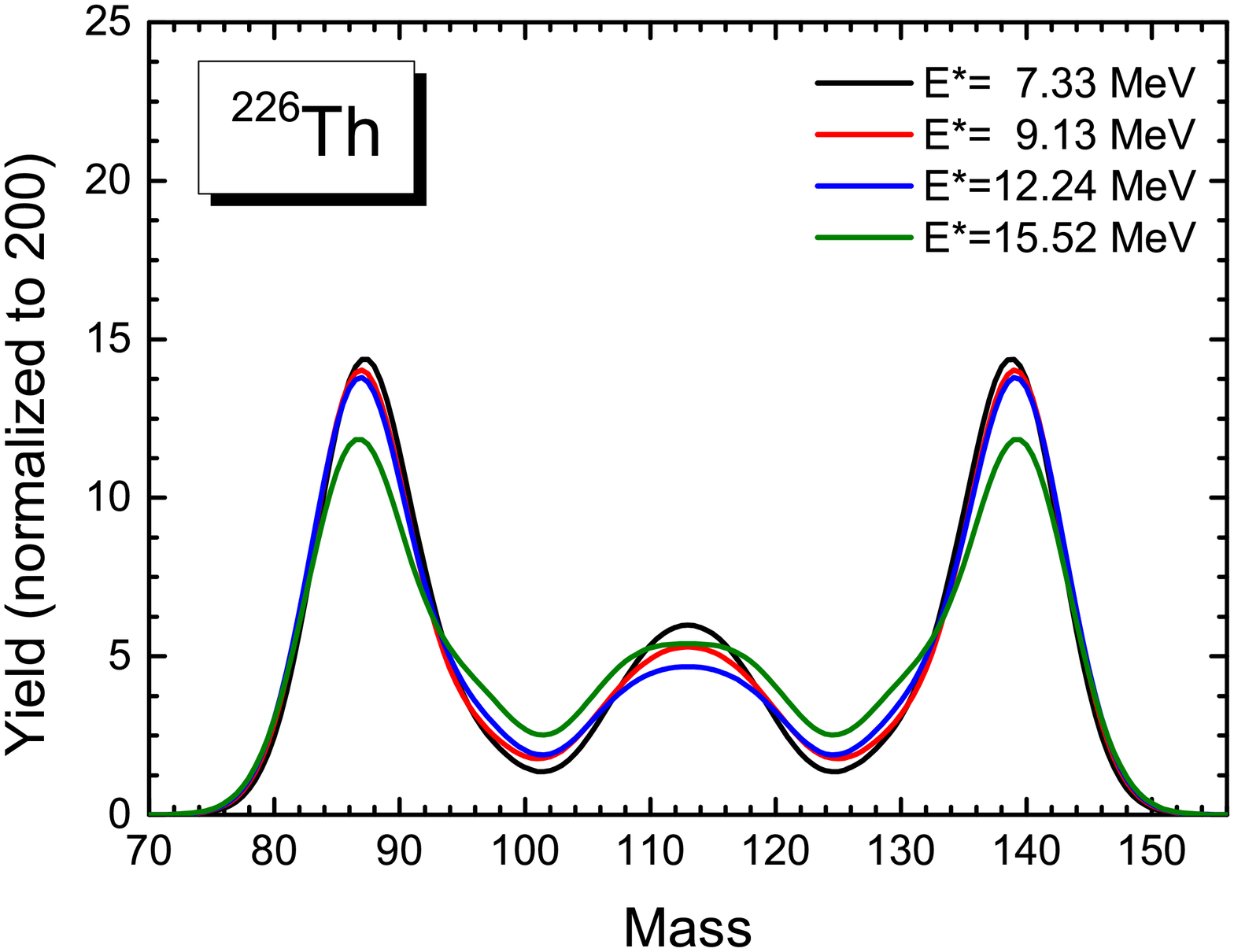}
\caption{\label{fig:MassEE}  (Color online)
Same as in the caption to Fig. \ref{fig:ChargeEE} but for the mass yield.}
\end{figure}

An interesting quantity to analyze is the energy dependence of the yields. In Figs.~\ref{fig:ChargeEE} and \ref{fig:MassEE} we show the charge and mass distributions of fission fragments for different excitation energies of the initial state, respectively. The original pairing strength is used. Both for the charge and mass distributions, one notices the transition from asymmetric to symmetric fission as the excitation energy increases. With the increase in energy the current can more directly enter the symmetric valley and, consequently, the asymmetric peaks are lowered while the symmetric peak gradually becomes wider. This result is consistent with the very recent prediction of the Metropolis walk method based on microscopic level densities \cite{Ward17}. At even higher energies, in the macroscopic limit, the yield distribution will eventually be dominated by symmetric fission.  However, we could not perform such a calculation here, because a boosted collective ground state does not represent a good choice for the initial state at very high excitation energy.

\section{\label{secIV} Summary and outlook}

The dynamics of induced fission of $^{226}$Th has been analyzed in a 
theoretical framework based on covariant energy density functionals and the corresponding
collective Hamiltonian, making use of a recently developed numerical implementation of
the time-dependent generator coordinate method plus Gaussian overlap approximation~\cite{Regnier16c}. 
The potential energy surface, mass tensor, scission line, and total kinetic energies have been calculated 
using the multidimensionally-constrained
relativistic mean-field model based on the energy density functional PC-PK1, and with pairing correlations 
taken into account in the BCS approximation. The fission process is described in a two-dimensional 
axially symmetric collective space ($\beta_{2}$, $\beta_{3}$). We note 
that the overall topography of the PES, the total kinetic energies for a particular pair of
fragments, and the general pattern of the scission line are consistent with previous studies
based on the Gogny effective interaction~\cite{Dubray08,Younes09}. 

The TDGCM+GOA calculation reproduces the main characteristics of the fission charge and mass
distributions, thus confirming the main conclusion of the analysis presented in Ref.~\cite{Regnier16}.  
By comparing the fission fragment yields for several values of the initial excitation energy, we have
found that increasing the latter leads to a lowering of asymmetric peaks and widens the
symmetric peak. 

The present study is based on the perturbative cranking approximation for the mass tensor.
It was shown, however, that this approximation underestimates the variation 
of mass parameters caused by level crossings~\cite{Baran2011_PRC84-054321}, and non-perturbative
cranking inertia can significantly modify spontaneous-fission paths and half-lives as compared 
to results obtained in the perturbative cranking approximation~\cite{Sadhukhan13,Jie15}.
The effects of non-perturbative cranking inertia on induced-fission mass distributions are presently not known, 
and this important topic will be the subject of our next study.

The importance of pairing correlations for the nuclear fission process has been demonstrated 
in numerous studies \cite{Negele78,Barranco90,Bertsch94,Bertsch97}. For instance, a recent investigation of fission 
dynamics of $^{240}$Pu within the real-time microscopic framework~\cite{Bulgac16} has shown that a number of 
shape and pairing modes are excited during the fission process. Studies of spontaneous fission~\cite{Sadhukhan16,Jie16} 
have shown the dramatic effect of the dynamical coupling between shape and pairing 
degrees of freedom on the calculated spontaneous fission life-times. In this study we have analyzed the influence
of ground-state pairing on the pre-neutron emission charge yields. The increase of static pairing correlations 
reduces the asymmetric peaks and enhances the symmetric peak in charge
yields distribution. Therefore a very interesting topic for future studies is dynamic pairing correlation in induced 
fission, possibly through the inclusion of pairing degrees of freedom
in the space of TDGCM+GOA collective coordinates.

\appendix
\section{\label{app-A} Axially deformed harmonic oscillator basis}

For a deformed axially symmetric shape the densities are invariant with respect to a rotation around the
symmetry axis, which is taken to be the $z$-axis here. It then turns out to be useful to work with cylindrical
coordinates
\begin{equation}
x=r_\perp\cos\varphi,\ \ \ \ y=r_\perp\sin\varphi, \ \ {\rm and}\ \ z.
\end{equation}
The single-nucleon Dirac spinors are expanded in terms of the eigenfunctions of a
deformed axially symmetric oscillator potential:
\begin{equation}
V(z,r_\perp)=\frac{1}{2}m\omega_z^2 z^2+\frac{1}{2}m\omega_r^2 r^2_\perp \;.
\end{equation}
Imposing volume conservation, the two oscillator frequencies $\hbar\omega_z$ and $\hbar\omega_r$
can be expressed in terms of a deformation parameter $\beta_0$:
\begin{eqnarray}
\hbar\omega_z            &=& \hbar\omega_0 \exp(-\sqrt{\frac{5}{4\pi}}\beta_0) \\
\hbar\omega_r  &=& \hbar\omega_0 \exp(\frac{1}{2}\sqrt{\frac{5}{4\pi}}\beta_0)
\end{eqnarray}
The corresponding oscillator length parameters are
\begin{equation}
b_z=\sqrt{\frac{\hbar}{m\omega_z}}\ \ \ {\rm and} \ \ b_r=\sqrt{\frac{\hbar}{m\omega_r}}
\end{equation}
The basis is now determined by the two constants $\hbar\omega_0$ and $\beta_0$.
The eigenfunctions of the deformed harmonic oscillator
potential are characterized by the set of quantum numbers
\begin{equation}
|\alpha\rangle=|n_z n_r m_l m_s\rangle
\end{equation}
where $m_l$ and $m_s$ are the components of the orbital angular momentum and spin
along the symmetry axis, respectively. The eigenvalue of $j_z$, which is a conserved quantity in this
case, is $\Omega=m_l+m_s$.
The eigenfunctions of the deformed harmonic oscillator can be explicitly written in the form:
\begin{equation}
\Phi_\alpha(z,r_\perp,\varphi, s, t)=\phi_{n_z}(z)\phi^{m_l}_{n_r}(r_\perp)\frac{1}{\sqrt{2\pi}}e^{im_l\varphi}
                                                                      \chi_{m_s}(s)\chi_{t_\alpha}(t)=\Phi_\alpha(r,s)\chi_{t_\alpha}(t)
\end{equation}
with
\begin{eqnarray}
\phi_{n_z}(z)                           &=& \frac{N_{n_z}}{\sqrt{b_z}}H_{n_z}(\zeta)e^{-\zeta^2/2} \\
\phi^{m_l}_{n_r}(r_\perp)  &=& \frac{N_{n_r}^{m_l}}{b_r}\sqrt{2}\eta^{m_l/2}L^{m_l}_{n_r}(\eta)e^{-\eta^2/2}
\end{eqnarray}
where $\zeta=z/b_z$ and $\eta=r_\perp^2/b^2_r$. $H_{n_z}(\zeta)$ and $L^{m_l}_{n_r}(\eta)$
are the Hermite polynomials and associated Laguerre polynomials, respectively. The normalization constants are given by
\begin{equation}
N_{n_z}=\frac{1}{\sqrt{\sqrt{\pi}2^{n_z}n_z!}} \ \ \ {\rm and} \ \ N^{m_l}_{n_r}=\sqrt{\frac{n_r!}{(n_r+m_l)!}} \;.
\end{equation}

\begin{figure}[htb]
\includegraphics[scale=0.35]{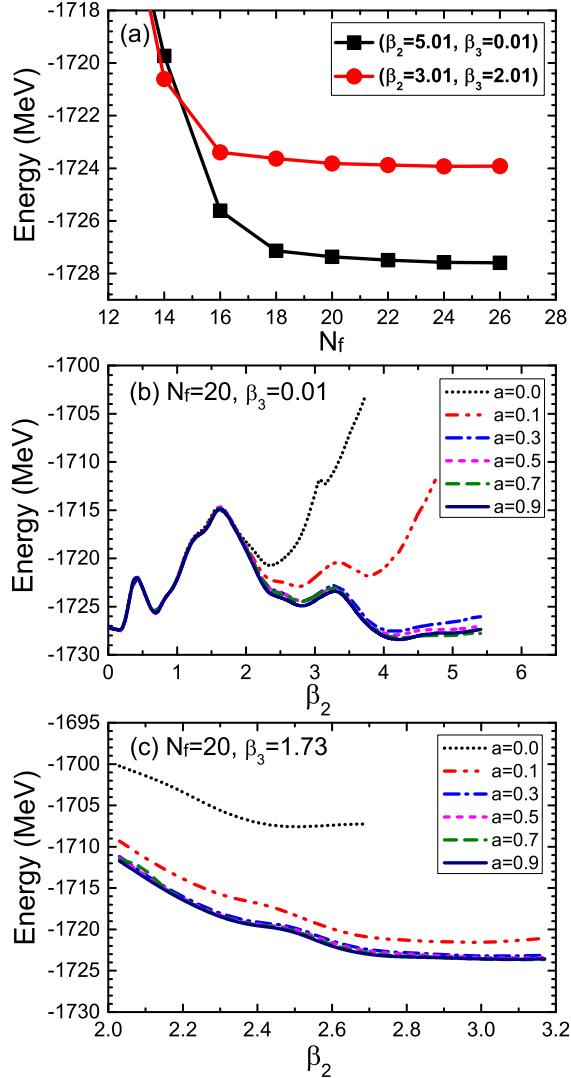}
\caption{\label{fig:shell}  (Color online) (a) Self-consistent RMF+BCS binding energy of $^{226}$Th for two extremely deformed configurations $(\beta_2, \beta_3)=(5.01, 0.01)$ and $(3.01, 2.01)$, as a function of the number of major shells of the axially deformed harmonic oscillator basis; (b) Self-consistent RMF+BCS binding energies of $^{226}$Th as functions of $\beta_2$ for different values of $a$ in Eq. (\ref{eq:beta0}), with $N_f=20$ and $\beta_3=0.01$; (c) Same as in the caption to panel (b) but for $\beta_3=1.73$.} 
\end{figure}

The Dirac spinor $\psi_i$, characterized by the quantum numbers $\Omega_i$ and isospin projection $t_i$,
can be expanded:
\begin{equation}
\label{eq:psi}
\psi_i(r,t)=\left(\begin{array}{c} f_i(r,s) \\ ig_i(r,s) \end{array} \right)\chi_{t_i}(t)
                =\left(\begin{array}{c} \sum\limits_\alpha^{\alpha_{\rm max}}f^i_\alpha\Phi_\alpha(r,s) \\
                   i \sum\limits_{\tilde\alpha}^{\tilde\alpha_{\rm max}}g^i_{\tilde\alpha}\Phi_{\tilde\alpha}(r,s)\end{array} \right)\chi_{t_i}(t)\;,
\end{equation}
and, of course, the summations in Eq. (\ref{eq:psi}) have to be truncated for a given number of shells $N_f$. Following the prescription of Ref. \cite{BNL14}, for the large component of the Dirac spinor all the states for which $[n_z/Q_z+(2n_r+|m_l|)/Q_r]\leq N_f$ are included in the expansion, where $Q_z={\rm max}(1, b_z/b_0)$ and $Q_r={\rm max}(1, b_r/b_0)$ are constants related to the oscillator lengths $b_0=\sqrt{\hbar /m\omega_0}$. To avoid the occurrence of spurious states, the expansion of the small component is truncated at $N_g=N_f+1$ major shells. In the present calculation, the parameters $\hbar\omega_0$ and $\beta_0$ are chosen as
\begin{equation}
\hbar\omega_0=41A^{-1/3} \ {\rm MeV},
\end{equation}
\begin{equation}
\label{eq:beta0}
\beta_0=\left\{
\begin{array}{ll}
0 & \ \ \  \rm{for} \ \ \beta_2<0; \\
a\beta_2 & \ \ \  \rm{for} \ \ 0\leq\beta_2\leq 1; \\
a\sqrt{\beta_2} & \ \ \  \rm{for}\ \  \beta_2>1.
\end{array}\right.
\end{equation}
For a given number of shells $N_f$ and following the procedure described above, a larger value of the parameter $a$ in Eq. (\ref{eq:beta0}) implies an increase in the number of basis states and, consequently, longer computing times. The convergence check for different values of $N_f$ and $a$ is illustrated in Fig. \ref{fig:shell}. It is found that the choice of the above parameters  largely mitigates basis truncation effects up to the scission point, where we estimate the error on the total energy to be $<1.0$ MeV for $N_f=20$ and $a=0.5$.

\begin{acknowledgements}
This work was supported in part by the NSFC under Grant No. 11475140,
the Croatian Science Foundation -- project ``Structure and Dynamics
of Exotic Femtosystems" (IP-2014-09-9159), the QuantiXLie Centre of Excellence,
the Chinese-Croatian project ``Microscopic Energy Density Functionals Theory for Nuclear Fission'',
and the NEWFELPRO project of the Ministry of Science, Croatia, co-financed
through the Marie Curie FP7-PEOPLE-2011-COFUND program.
\end{acknowledgements}


\end{document}